\documentclass[10pt]{article}
\usepackage{graphicx}
\begin{document}

\begin{center}
{\bfseries Laser acceleration of ion beams \footnote{{\small Talk
at the Helmholtz International Summer School "Dense Matter In
Heavy Ion Collisions and Astrophysics", JINR, Dubna, August 21 -
September 1, 2006.}}}

\vskip 4mm

I.A. Egorova$^{1,3}$, A.V. Filatov$^1$, A.V. Prozorkevich$^1$,
S.A. Smolyansky$^1$ \\
D.B. Blaschke$^{2,3,4}$, M. Chubaryan$^{3}$
\vskip 3mm

{\small $^1${\it Saratov State University, 410026 Saratov,
Russia}}\\
{\small$^2${\it Institute for Theoretical Physics, University of Wroclaw,
50-204 Wroclaw, Poland}}\\
{\small$^3${\it Joint Institute for
Nuclear Research, 141980 Dubna, Russia}}\\
{\small$^4${\it Institute of Physics, Rostock University,
D-18051 Rostock, Germany}}
\end{center}

\vskip 3mm

\centerline{\bf Abstract}

We consider methods of charged particle acceleration by means of
high-intensity lasers. As an application we discuss a laser booster
for heavy ion beams provided, e.g. by the Dubna nuclotron.
Simple estimates show that a cascade of crossed laser beams would be necessary
to provide additional acceleration to gold ions of the order of GeV/nucleon.

\vskip 5mm

The idea of particle acceleration by means of laser fields was
first considered in \cite{ESCH,Ask}. Recently, due to the fast
development of ``table-top'' high-intensity lasers with
intensities up to $10^{22}$ W/cm$^2$ \cite{Paris}, these systems
have become interesting for the development of booster systems for
the upgrade of existing accelerators which operate just below the
threshold for interesting physics. In the case of the  Dubna
nuclotron an additional acceleration by a few GeV/nucleon would
allow studies of the onset of quark matter formation in heavy-ion
collisions. Such studies require a fine-tuning in beam energies
(energy scan), which be possible with the laser booster provided
the required range of acceleration can be reached.

We start by considering the light pressure action on a
beam of charged particles.
When absorbing a photon the nucleus makes a transition to an excited state
and receives an additional velocity in the direction of the photon propagation.
After that, the excited nucleus can spontaneously emit a photon and receives
some additional momentum.
As a result of multiple scattering, the momentum of the nucleus gets
incremented along the laser beam.
In the oscillator model of nuclei one obtains the following estimate for the
light pressure \cite{Ask}:
\begin{equation}\label{P}
P=\frac{2\alpha^2}{3\lambda^4}{\mathcal{E}}^2,
\end{equation}
where $\mathcal{E} = \sqrt{< \mathbf{E}^2>} $ is the mean electric
field strength, $\alpha(\omega)$ is the polarizability  of the
given nucleus
\begin{equation}\label{alpha}
\alpha(\omega)\cong-\frac{e_i^2}{m_i}\sum_nk_n[\omega^2-\omega_n^2]^{-1},
\end{equation}
$\hbar\omega_n$ is the n-th energy level of the nucleus, and $k_n$ is the
corresponding oscillator strength taking into account the the specifics of the
nucleus.

The application of the light pressure mechanism to the problem of ion
acceleration encounters numerous obstacles. First, it is the
saturation effect which is associated with the time delay of the
vacation of the excited state of nuclei. Second, there is the
required condition of resonance between the laser field frequency and
nuclear excitation energy.
When the nucleus receives additional energy due to the action of the laser
pulse, the condition of resonance brakes because of the Doppler effect and
therefore the acceleration force will decrease.
The electromagnetic wave
frequency  in the proper reference frame $\omega$
and in the laboratory one $\omega'$ are related by by
\begin{equation}\label{e1}
\omega^{\,\prime} = \frac{\varepsilon^*}{\hbar} (1+v/c) \gamma, \qquad \qquad
\gamma= 1/\sqrt{1 - v^2/c^2},
\end{equation}
where $\varepsilon^* = \hbar \omega $ is the transition energy to
an excited state. For the Dubna nuclotron $\gamma \simeq 4$, that
corresponds to the ultra-relativistic case and so $\omega' \simeq
2\gamma \varepsilon^*\hbar^{-1}$.
If the energy $\varepsilon^*$ has the order of some tens of keV
(for $^{179}$Au we have $\varepsilon^* = 72$ keV), the X-ray laser
should have a frequency of the order of 1 MeV.
This exceeds considerably the possibilities of modern X-ray lasers
\cite{Ring}, for which the typical values of frequencies are less
than 10\% of the electron mass.

In case of optical lasers, when the frequency is much smaller than
the lowest nuclear excitation level, there are other ways of
charged particle acceleration by means of the action of gradient
forces caused by the interaction between ions and the envelope of
the laser electrical field. Let us consider a harmonic field $
\mathbf{E}(\textbf{r},t)= \mathbf{E}(\textbf{r})\exp({i\omega
t})$. The corresponding gradient force is \cite{Ask,AskM,Gaponov}
\begin{equation}\label{force}
\mathbf{F} = \alpha \nabla {\mathcal{E}}^2 .
\end{equation}
Under the action of such force, ions move aside to decrease their
potential energy and get localized in the knots of the wave.
In each point the gradient  force has transverse and longitudinal
components but the average of the longitudinal component vanishes.
Under the action of the gradient force particles are
localized in potential wells.
As was shown in \cite{Gaponov2} using special fluctuations with various
frequencies it is possible to receive a potential profile varying in time.
In particular, it is possible to create conditions for accelerated motion of
potential wells with particles localized in them.
However, there is a grave disadvantage of this method.
That is the fact that the accelerated particle oscillates with the frequency
of the external field in the local reference frame.
The efficiency of such a mechanism is less than that of a linear accelerator.
In real laser beams due to the non-uniform radial distribution of intensity
the gradient force can expel particles from the laser field.
As shown in \cite{Ask62}, this effect depends on the relation of the frequency
of the external field to the eigenfrequency of the particle oscillation.

Another way to transfer additional energy to charged particles is
to  construct special geometrical schemes of laser beams.
A scheme to accelerate electrons by means of two crossed laser beams
was proposed in the works \cite{Haaland}-\cite{Salamin2000}.
The basic idea is to send the electron or ion through the crossing
point of two laser beams at an angle $\theta$ with respect to each
beam direction \cite{Salamin2003}.
Such a geometry allows to create some longitudinally pulling electric field
\begin{eqnarray}
  E_z = -2E_0\, g(\eta)\, \sin{\theta}\, \cos{\eta}, ~~ 
\qquad \quad \eta =\omega\, [\, t-(z/c)\cos{\theta} ],
\end{eqnarray}
where $g(\eta)$ is some envelope function.
The ion energy gain, as a result of the interaction with $\eta/2\pi$ cycles of
the accelerating field, may be defined by
\begin{eqnarray}\label{s1}
W(\eta) &=& m c^2 \left[\frac{s \cos{\theta} + \sqrt{s^2 +
\sin^2{\theta}}}{\sin^2{\eta}} -\gamma_0 \right], \nonumber\\ s
&=& \gamma_0 (\beta_0 - \cos{\theta}) + 2 q f(\eta) \sin{\theta},
\end{eqnarray}
where $\gamma_0$ is initial value of the Lorentz-factor for $\beta_0=v_0/c$,
\begin{eqnarray}\label{s3}
f(\eta) &=& \int\limits_{\eta_0}^{\eta} g(\eta') \cos{\eta'}
d\eta',
\end{eqnarray}
$q = ZeE_0/(M c \omega)$ for an ion with electrical charge $Ze$ and a mass $M$.

\begin{figure}[h]
\centering
\includegraphics[width=0.47\textwidth,height=50mm]{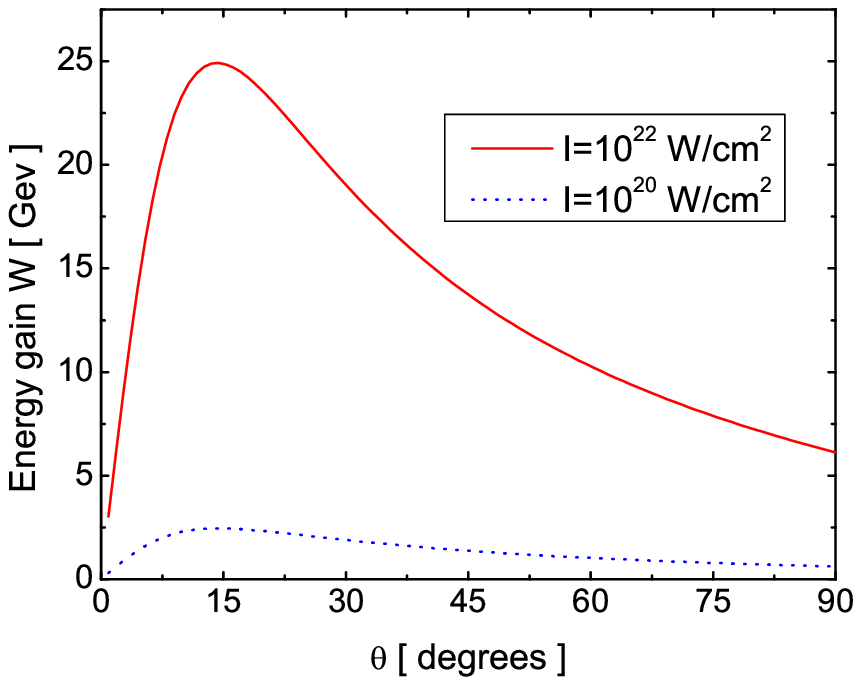}
\hfill
\includegraphics[width=0.47\textwidth,height=51mm]{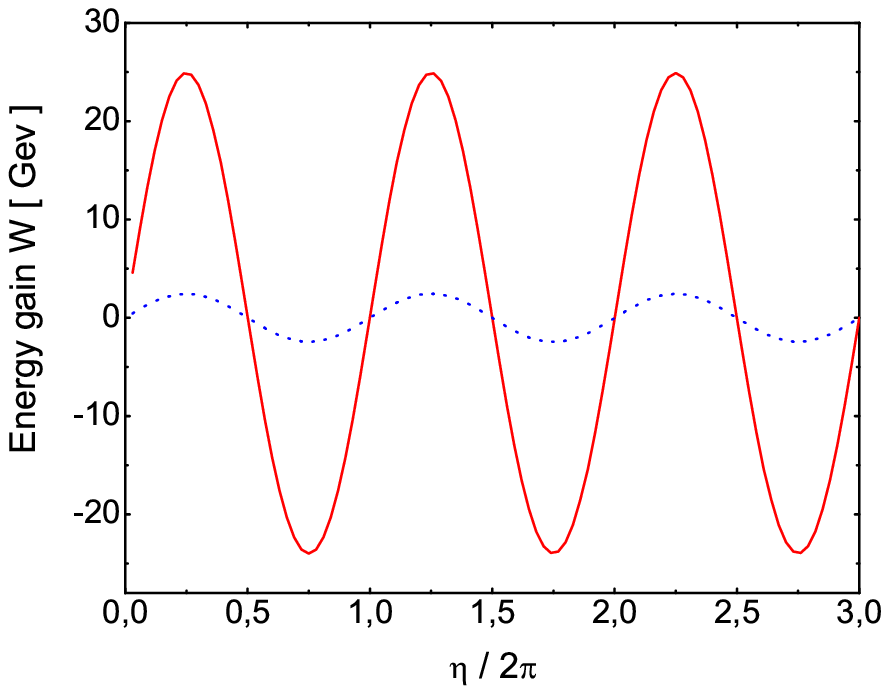}
\caption{\small{The energy gain of fully ionized gold ions Au$^{79+}$ vs.
the crossing half-angle $\theta$ for $\eta =\pi/2$ (left) and also
vs. the number of field cycles for
$\theta$=$\theta_{\rm max}$=14$^\circ$(right), $\gamma_0$=4, g($\eta$)=1.}}
\label{fig1}
\end{figure}

The size of the energy increment depends on the crossing angle.
An optimal angle $\theta_{\rm max}$ exists which corresponds to the maximum
energy gain for a given set of laser parameters and particle beam
initial conditions.
Fig.\ref{fig1} shows, that if the ion-field interaction would be terminated
in the neighborhood of any one of the
points corresponding to $\eta =(2N+1/2)\pi$, where $N$ is an integer number,
the ion would escape with maximum energy gain.
The energy gains tend to get canceled by the energy losses for interaction
with an even number of field cycles.
For the purpose of acceleration, the assumption is that the ion may be
ejected from the region of interaction while it still retains part or all of
the energy gained.
The calculation shows that the optimal half-crossing angle
for Au$^{79+}$ is $\theta_{\rm max} \approx 14^\circ$ for
$g(\eta)=1$  and $\theta_{\rm max}\approx 20^\circ$ for the $g(\eta)=
\sin^2(\eta/10)$ envelope.
The maximal energy gain reaches 25 GeV per ion (140 MeV/nucleon) for a laser
intensity of the order of $10^{22}$ W/cm$^2$.
The value of the optimal angle allows the realization of a cascade of
laser - ion beam interactions  by means of multiple reflection of the laser
beams from a sequence of mirrors positioned symmetric to the beam axis.

The initial condition  used here are somewhat artificial. We have
simply stated that the ion is born at the origin of coordinates at
$t=0$ inside the plane wave which, by definition, has an infinite
extension in both space and time. Fig.2 shows that such an ion
tends to gain more energy from the field if it starts off at a
higher speed.
The gain exhibits saturation with increasing injection energy at about
$\gamma_0\approx 20$ for the parameters used.
The radiative losses of the accelerated ion can
be evaluated by the relativistic version of the Larmor formula
\cite{Jackson} and become negligible.

\begin{figure}[h]
\centering
\includegraphics[width=0.47\textwidth,height=50mm]{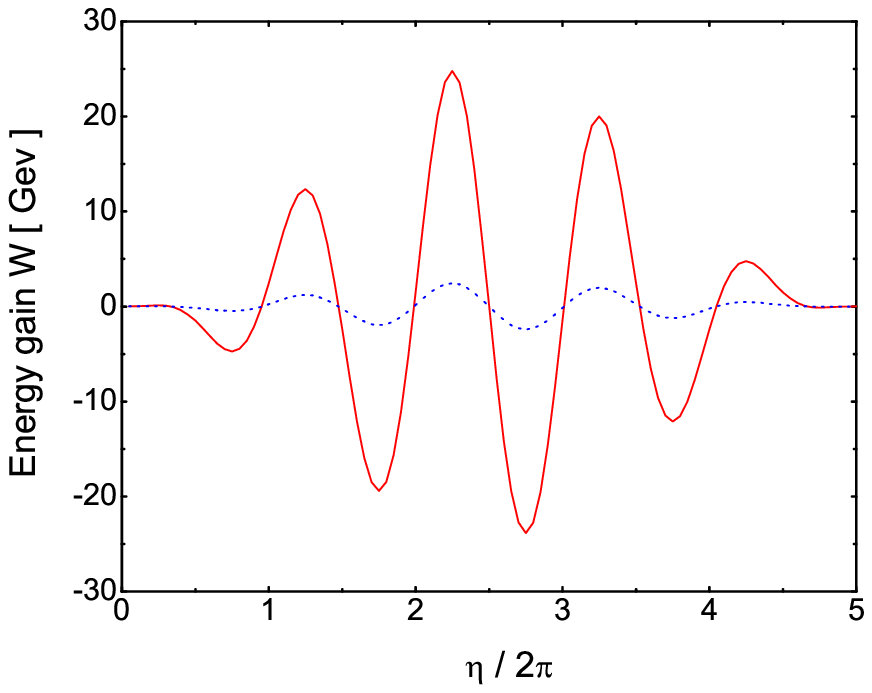}
\hspace{5mm}
\includegraphics[width=0.47\textwidth,height=50mm]{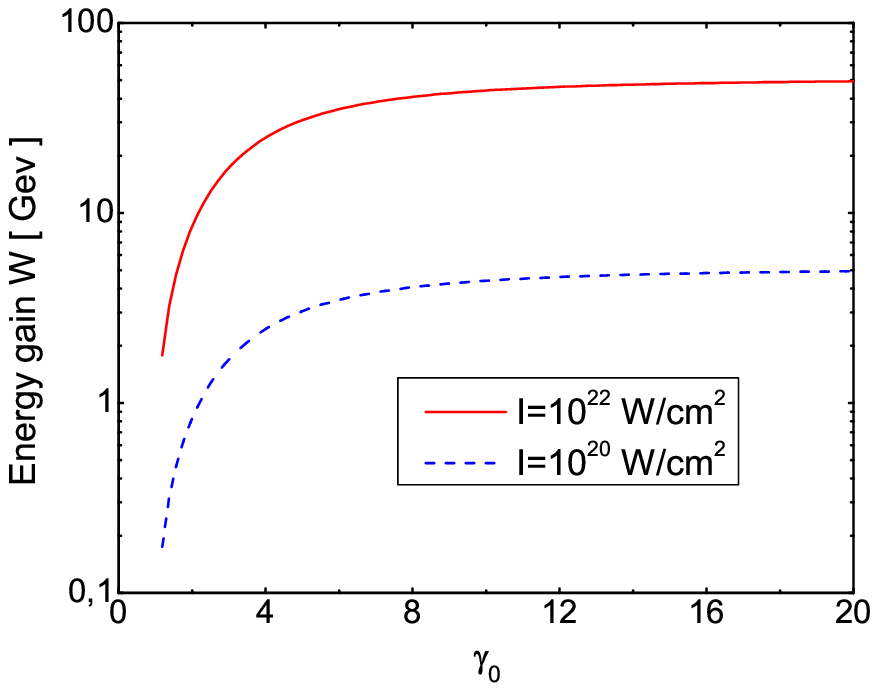}
\caption{\small{The energy gain of Au$^{79+}$ vs. the number of
accelerating field cycles for
$\theta$=$\theta_{max}\approx$20$^\circ$ (left) and also vs the
initial Lorentz factor $\gamma_0$ for $\eta =9\pi/2$ (right),
g($\eta$)= sin$^2$($\eta/10$). The remaining field parameters used
are the same as in Fig.1.}}\label{fig2}
\end{figure}

Our preliminary evaluation shows that the most promising method
for the laser acceleration of ions is based on the crossed laser
beam scheme. The light pressure mechanism in not relevant due to
insufficient frequency of the laser field even for modern X-ray
lasers. The action of gradient forces on laser beam is also less
effective than the usual linear accelerator scheme. We have
disregarded here some effects which are important for the
practical realization of the laser booster as, for example, the
presence of residual gas in the channel of the ion beam.

\end{document}